\documentclass[prd,aps,twocolumn,a4paper,floatfix]{revtex4-1}
\usepackage{hyperref}

\def\p{\partial}
\newcommand{\As}{A_\star}

\usepackage{graphicx,psfrag}
\usepackage{mathrsfs}
\usepackage{amsmath,amsfonts,amssymb,amsthm}
\usepackage{url}

\begin{document}

\title{Evolutions of centered Brill waves with a pseudospectral 
method}

\author{David Hilditch$^{1,2}$}
\author{Andreas Weyhausen$^2$}
\author{Bernd Br\"ugmann$^2$}

\affiliation{${}^1$CENTRA, University of Lisbon, 1049 Lisboa, Portugal.\\
${}^2$Friedrich-Schiller-Universit\"at Jena, 07743 Jena, Germany.
}

\begin{abstract}
The pseudospectral code \texttt{bamps} is used to evolve
axisymmetric gravitational waves. We consider a one-parameter
family of Brill wave initial data, taking the seed function and
strength parameter of Holz~et.~al. A careful comparison is made
to earlier work. Our results are mostly in agreement with the
literature, but we do find that some amplitudes reported elsewhere
as subcritical evolve to form apparent horizons. Related to this
point we find that by altering the slicing condition, the position
of the peak of the Kretschmann scalar in these supercritical data
can be controlled so that it appears away from the symmetry axis
before the method fails, demonstrating that such behavior is at
least partially a coordinate effect. We are able to tune the
strength parameter to an interval of range~$1-A_\star/A\simeq10^{-6}$
around the onset of apparent horizon formation. In this regime we
find that large spikes appear in the Kretschmann scalar on the
symmetry axis but away from the origin. From the supercritical
side disjoint apparent horizons form around these spikes. On the
subcritical side, down to this range, evidence of power-law scaling
of the Kretschmann scalar is not conclusive, but the data can be fitted
as a power-law with periodic wiggle.
\end{abstract}


\maketitle

\section{Introduction}\label{section:Introduction}

Motivated partially by the findings of~\cite{HilBauWey13}, in
particular by our difficulties in evolving gravitational wave data
close to the critical threshold of black hole formation with the
moving puncture gauge, we turned to an alternative formulation and a
more accurate numerical method. We implemented the generalized
harmonic formulation in a pseudospectral code, \texttt{bamps}, which
was recently described in detail
in~\cite{HilWeyBru15,BugDieBer15,HilHarBug16}.
Presently we use this tool to evolve Brill wave initial
data~\cite{Bri59} in the form most often treated numerically.
Primarily we choose such data for ease of comparison with the
literature, but additionally since it is axisymmetric it allows
us to run the code most efficiently. Ultimately we hope to obtain
a proper understanding of, and a robust numerical method for
gravitational waves close to the threshold of black hole formation.
This study, like~\cite{HilHarBug16}, is another step in that
direction.

The key results in the literature on critical collapse of
gravitational waves are those of Abrahams and
Evans~\cite{AbrEva93,AbrEva94}, who considered one-parameter families
of Teukolsky wave initial data in axisymmetry, and found that the
resulting black hole mass scales as a power law in a neighborhood of
the critical threshold. They also found more tentative evidence for
`echoing', or periodicity in spacetime scale, of the solution.
Primarily because of its simplicity, most subsequent studies have focused on
Brill wave initial data. In evolutions of Brill waves Sorkin~\cite{Sor10}
found evidence for {\it another} critical solution in which the peak of the
curvature appears on concentric rings around the symmetry axis. This study
employed the generalized harmonic formulation in axisymmetry, and so is the
natural starting point for us. For a detailed discussion of critical phenomena
in gravitational collapse see~\cite{GunGar07}.

The paper is structured as follows. In section~\ref{section:Setup}, 
we summarize the main features of the~\texttt{bamps} code and the continuum 
equations. In section~\ref{section:Brill_pos} we present our evolutions.
Afterwards we conclude in section~\ref{section:Conclusions}.

\section{Setup}
\label{section:Setup}

The~\texttt{bamps} code uses a pseudospectral method to evolve 
a first order formulation~\cite{LinSchKid05} of the generalized 
harmonic gauge formulation, with
\begin{align}
\p_t\alpha&=-\alpha^2 K+\eta_L\alpha^2\log\left(\tfrac{\gamma^{p/2}}
{\alpha}\right)+\beta^i\p_i\alpha\,,\nonumber\\
\p_t\beta^i&=\alpha^2\,{}^{\textrm{\tiny{(3)}}}\Gamma^i-\alpha\,\p^i\alpha
-\eta_S\beta^i+\beta^j\p_j\beta^i\,,\label{eqn:3+1_ghg}
\end{align}
in the standard~$3+1$ notation. We parametrize the free scalars
by~$\eta_L=\bar{\eta}_L\alpha^q$ and~$\eta_S=\bar{\eta}_S\alpha^r$,
with~$\bar{\eta}_L,\bar{\eta}_S,p,q$ and~$r=0$ constant. We employ radiation
controlling constraint preserving boundary conditions 
like those described in~\cite{Rin06a,RinLinSch07}, 
imposed via the Bj\o rhus method~\cite{Bjo95}, but modified to 
minimize reflections caused by the use of constraint damping, which
can otherwise cause the code to crash with our gauge conditions.
The numerical method is similar to that of SpEC~\cite{spec}, employing 
many subpatches across which data is communicated using a 
penalty approach. For our grids we presently take either cubed-sphere 
or cubed shells~\cite{RonIacPao96,Tho04}. We discretize in 
space using Chebyschev polynomials, filtering the highest order 
coefficients~\cite{SziLinSch09}. Although \texttt{bamps} is fully 3d, 
to evolve efficiently in axisymmetry we have implemented the 
analytic-cartoon method~\cite{Pre05,AlcBraBru99a}, in which 
all angular spatial derivatives are evaluated using the Killing vector. 
The resulting regularity conditions are imposed on axis and at 
the origin. The code is parallelized with MPI at the subpatch level,
with one or more subpatches per core, resulting in excellent strong-scaling
for up to several thousand cores. Presently we evolve Brill wave initial 
data~\cite{Bri59,Epp77}, in which the spatial metric takes 
the form,
\begin{align}
\textrm{d}l^2&=\gamma_{ij}\textrm{d}x^i\textrm{d}x^j
=\Psi^4\big[e^{2q}(\textrm{d}\rho^2+\textrm{d}z^2)
+\rho^2\textrm{d}\phi^2\big]\,,\label{eqn:Brill_ansatz}
\end{align}
and the extrinsic curvature vanishes. We choose always the seed 
function~$q$ of~\cite{HolMilWak93}, 
\begin{align}
q(\rho,z)&=A\rho^2e^{-[(\rho-\rho_0)^2/\sigma_\rho^2+(z-z_0)^2/\sigma_z^2]}\,,
\label{eqn:brill_seed}
\end{align}
with~$A>0$,~$\sigma_\rho=\sigma_z=1$ and~$\rho_0=z_0=0$. We call this 
data a centered geometrically prolate Brill wave. To build initial data
we use the solver of~\cite{DieBru13}. Our main tool for classifying 
spacetimes as supercritical is the axisymmetric apparent horizon 
finder~\texttt{AHloc}, which we run in postprocessing. A detailed 
description of the setup can be found in~\cite{HilWeyBru15,Wey14}.

\section{Centered Geometrically prolate Brill waves}
\label{section:Brill_pos}

In this section we present our evolutions of centered geometrically 
prolate Brill waves. The first numerical evolutions of Brill waves 
that we are aware of were presented in~\cite{Epp79}. Since then 
Brill wave initial data have been considered multiple times in the 
numerical relativity
literature~\cite{AlcAllBru99,GarDun00,Rin06,San06,HilBauWey13}.
Therefore to ensure that~\texttt{bamps} is performing properly we
start with a detailed comparison of the evolutions performed with the
seed function~\eqref{eqn:brill_seed} away from criticality. We then
compare with the results of~\cite{Sor10} before going towards
criticality.

\subsection{Review and comparison with earlier work}
\label{subsection:comparison}

\paragraph*{Alcubierre et. al.:} In~\cite{AlcAllBru99} Brill 
waves were evolved numerically for the first time in 3d. The 
BSSNOK formulation was used in combination with maximal slicing 
and vanishing shift. Using the given data~\eqref{eqn:brill_seed}
with~$\rho_0=z_0=0$, $\sigma_\rho=\sigma_\rho=1$, which we consider
throughout, it was found that the critical point lies
between~$A=4$ and~$A=6$, and furthermore that this finding could be
refined to~$A=4.85\pm0.15$, although the data for this latter claim
were never presented. Supercriticality was diagnosed by finding 
an apparent horizon, which occurred for the~$A=6$ data 
at~$t=7.7$. In~\cite{GunGar06} it was shown that BSSNOK
combined with this gauge choice results in an ill-posed PDE 
system, meaning that this approach should be either abandoned 
or modified if we are to achieve accurate results that converge
to the continuum solution as resolution is increased.

\paragraph*{Garfinkle and Duncan:} In~\cite{GarDun00} it was 
found, evolving Brill wave initial data with~$q$ as 
in~\eqref{eqn:brill_seed}, again taking~$\rho_0=z_0=0$,
$\sigma_\rho=\sigma_\rho=1$, that the critical amplitude~$\As$ lies
between~$A=4$ and~$A=6$. The data was classified either by evolving
until the spacetime was close to flat and subsequent collapse seemed
implausible, or by explicitly finding an apparent horizon. The formulation
employed was explicitly axisymmetric, and consisted of a mixed
elliptic-hyperbolic system with maximal slicing~$K=0$, well-posedness of
which, to the best of our knowledge, has not been studied. We agree with
the findings of both~\cite{AlcAllBru99} and~\cite{GarDun00}. Because our
method employs a different gauge however, it is difficult to make a
side-by-side comparison beyond classifying the spacetimes as sub or
supercritical. The effect of changing the shape of the initial data
parameters~$\sigma_\rho$ and~$\sigma_z$ was also studied in~\cite{GarDun00},
but we have not yet followed up on this.

\begin{figure*}[t]
\centering
\includegraphics[width=\columnwidth]{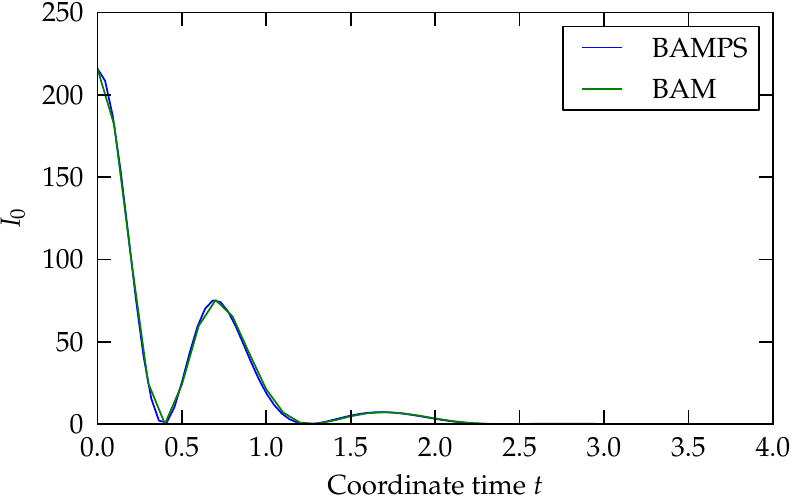}
\includegraphics[width=\columnwidth]{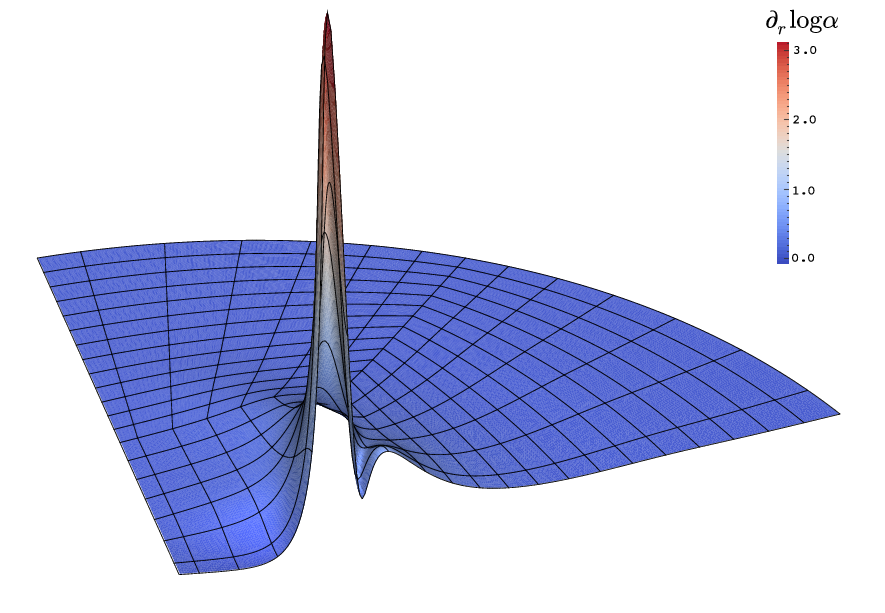}\\
\includegraphics[width=\columnwidth]{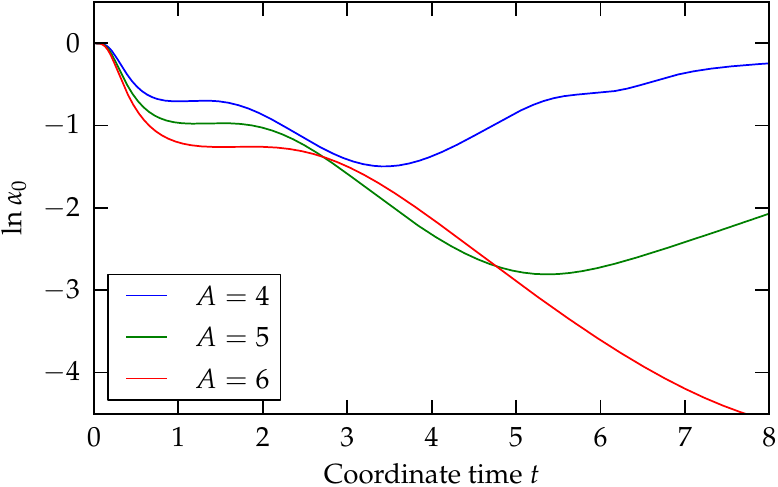}
\includegraphics[width=\columnwidth]{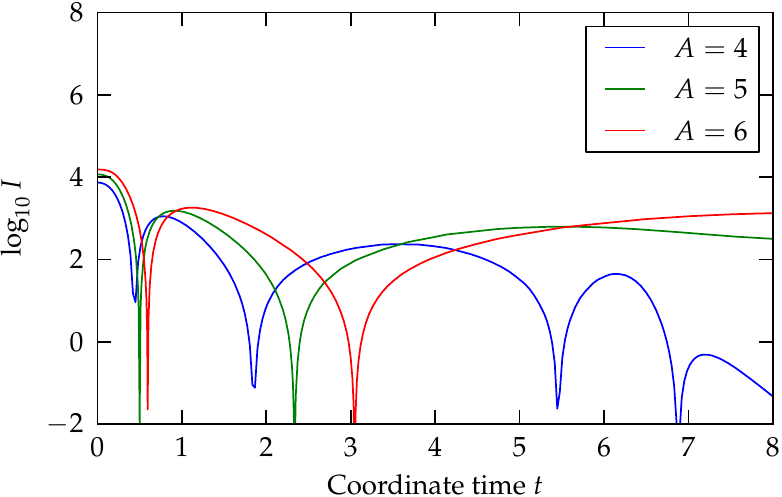}
\caption[]{On the upper left we show the central value of the Kretschmann 
scalar~$I_0$ in a centered Brill wave~$A=1$ evolution, with BAM and~\texttt{bamps}.
BAM was used with moving puncture coordinates (data taken from 
in~\cite{HilBauWey13}) and~\texttt{bamps} was run with pure harmonic slicing. 
This figure is meant to be directly compared to Fig.~9.2 of~\cite{Rin06}. 
The fact that we employ a different spatial gauge does not matter 
here because harmonic lapse is a pure slicing condition, so values 
of the lapse function can be compared one-to-one at the symmetry 
axis. On the upper right we show a snapshot of~$\p_r \alpha$ in the~$xz$ 
plane at~$t\approx1.72$, which should be compared with~Fig.~$9.3$ 
of~\cite{Rin06}. Although the spatial coordinates differ, there 
is an obvious qualitative agreement. In the bottom two panels the comparison 
with the work of Rinne~\cite{Rin06} continues. In these runs we evolve a brill 
wave using harmonic lapse and~$\eta_S=6.0$ for the shift. On the left we show 
the logarithm of the central lapse over time for~$A=4, 5$ and~$6$. These
results should be compared with Fig.~$9.9$ in~\cite{Rin06}. The line
for~$A=4$ agrees quite well. The others disagree. On the right we show
the central value of the Kretschmann scalar.
\label{fig:Rinne_A1_Kretsch_Lapse_Kretsch}
}
\end{figure*}

\begin{figure}[t]
\centering
\includegraphics[width=\columnwidth]{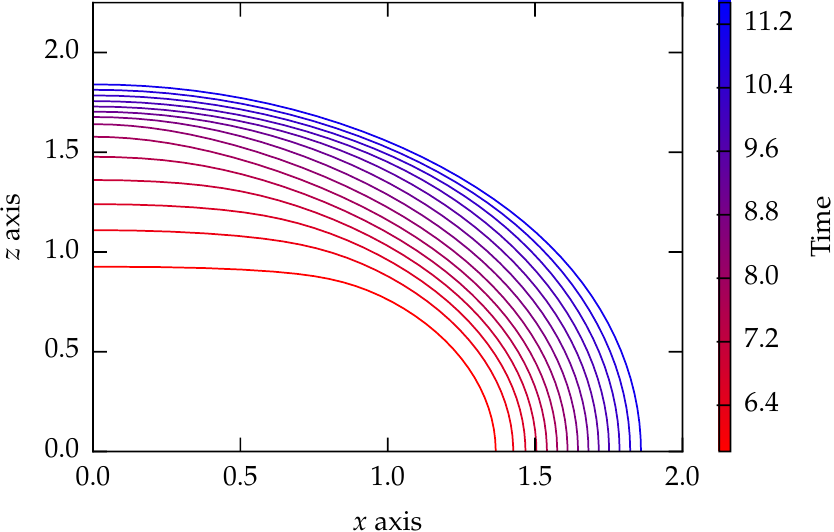}
\caption[]{
Here we plot the development of the apparent horizon for Brill waves with
amplitude~$A=5$, evolved with pure harmonic slicing and damped harmonic shift.
The horizon is first found at~$t=5.82$, with area~$A_H=15.7$. At the end of the
evolution at~$t=11.9$ the area has increased to~$A_H=16.7$. Similar results are
obtained in the~$A=6$ evolution. The initial area at~$t=2.1$ 
is~$A_H=33.7$ and the final is~$A_H=40.6$ at $t=8.24$.
\label{fig:Rinne_Horizon}
}
\end{figure}

\paragraph*{Rinne:} In the PhD thesis~\cite{Rin06} evolutions 
of centered geometrically prolate Brill waves were presented with
a free-evolution and a partially constrained scheme, both in
explicit axisymmetry. We focus on the free-evolution scheme, the~Z$(2+1)+1$ 
formulation, since that is where we are able to make the clearest 
comparisons. In that case harmonic slicing was taken with vanishing 
shift. This choice is convenient for our comparison because although 
we can not choose the same shift condition, harmonic lapse is a 
pure slicing condition, which means that we should obtain the same 
foliation of the same spacetime (starting from the same initial lapse)
albeit with different spatial coordinates. Since there is a preferred
observer, namely that at the origin, we can compare quantities explicitly
there. Fortunately the work~\cite{Rin06} contains several plots along
this worldline. In the upper left panel of
Fig.~\ref{fig:Rinne_A1_Kretsch_Lapse_Kretsch} we plot the Kretschmann
scalar
\begin{align}
I=R_{abcd}R^{abcd}\,,
\end{align}
at the origin as a function of time for~$A=1$ centered Brill wave data, 
which should be compared with Fig.~9.2 of~\cite{Rin06}. In this test 
we evolved with harmonic slicing~$\eta_L=0$ and the damped harmonic 
shift~$\eta_S=6$ and otherwise our standard setup. The agreement,
at least by eye, is extremely good. Taking~$A=4$ it was found that
with sufficient resolution a sharp feature in the gradient of the lapse
could be resolved. It was found that the data was, in agreement
with~\cite{GarDun00}, subcritical. We see the same result. In the upper right
panel of Fig.~\ref{fig:Rinne_A1_Kretsch_Lapse_Kretsch} we
show~$\p_\rho\ln\alpha$ at~$t=1.72$. This is the time at which Rinne finds
the largest peak in this quantity. A similar plot is Fig.~$9.3$
in~\cite{Rin06}, which can not be directly compared because of the differing
spatial coordinates, although the qualitative agreement is very clear. We 
find that the largest peak appears at around~$t=2.08$, but the 
magnitudes in~$\p_\rho\ln\alpha$ differ by less than~$5\%$ across 
these times. In the lower two panels of
Fig.~\ref{fig:Rinne_A1_Kretsch_Lapse_Kretsch}, to be compared with
Fig.~$9.9$ of~\cite{Rin06}, the left panel shows the logarithm of
the lapse at the origin as a function of coordinate time for
amplitudes~$A=4,5,6$. In the right hand panel we plot the value of the 
Kretschmann scalar at the origin. At lower resolutions we find that the 
Kretschmann scalar exhibits high-frequency oscillatory behavior, but 
that these wiggles converge away rapidly. The most challenging data 
evolved in Rinne's experiments was the~$A=5$ wave, for which sub and 
supercriticality was not discerned using the free-evolution algorithm, 
partially because the data was still oscillatory at the time the method 
failed at around~$t=6$. All of the different resolutions we tried 
with~$A=5$ in this suite of tests crashed at coordinate time~$t\simeq12$. 
We ran our apparent horizon finder on this data, the result of which is 
plotted in Fig.~\ref{fig:Rinne_Horizon}. We find the apparent horizon
first at~$t\simeq6.2$ and thereafter until the evolution fails.
Rinne correctly concluded that the critical amplitude lies below~$A=6$,
although no apparent horizon could be found in his data. Instead
his classification was made by observing that the Kretschmann scalar
was blowing-up. This diagnostic is flawed because as one approaches
the critical point we expect to generate arbitrarily large curvature
scalars even in subcritical data. In the absence of an apparent horizon
or event horizon however, other diagnostics may be similarly flawed,
and the Kretschmann scalar is at least a spacetime scalar, so one might prefer
it as a diagnostic to `collapse of the lapse'~\cite{Alc08} which is clearly
gauge dependent. Running our apparent horizon finder on the~$A=6$ data we find
an apparent horizon at~$t\simeq2$ and later.

\paragraph*{Previous studies with BAM:} In $2005$~\cite{San06} the 
BAM finite differencing code~\cite{HilBerThi12,ThiBerBru11,BruGonHan06} 
was used to evolve centered geometrically prolate Brill waves
with the BSSNOK formulation combined with several different 
gauge conditions, maximal slicing, harmonic slicing, and the
moving puncture gauge condition. The main complications were
reported to be constraint violation, which was likely caused 
by lack of resolution, and which did vary significantly from
one gauge to another. To understand how much further, 
if at all, standard modern numerical relativity methods can 
go beyond those previously discussed, recently in~\cite{HilBauWey13} 
the BAM code was once again used, this time alongside the code
of~\cite{BauMonCor12} with the focus purely 
on the moving puncture gauge. Starting with~$A=1$ (that is, weak) data, it 
was found that the lapse initially decreased, but rapidly returned 
back to unity; we find qualitatively the same behavior despite 
the different gauge used in~\texttt{bamps}, although the lapse 
function decreases by a smaller amount in the new data. The Kretschmann
scalar decreases from a maximum of about 216 at~$t=0$ to zero and reaches
a second maximum at~$t=0.7$. The maximum value of the Kretschmann scalar
in the domain immediately decreases and never grows beyond the initial
value as the wave propagates away. In the upper left panel of 
Fig.~\ref{fig:Rinne_A1_Kretsch_Lapse_Kretsch} we also plot the central value 
of the Kretschmann scalar obtained in this BAM experiment. Since the time
coordinates used differ, the horizontal axes would be different, but the
values of the maxima should be the same. The BAM value is about~$I_0\sim75$
which agrees extremely well with the~\texttt{bamps} experiment. The
Hamiltonian constraint violation in this particular BAM run is of the
order~$10^{-3}$ at the time of the second local in time maximum, whereas
the roughly analogous~$F_t$ constraint inside~\texttt{bamps} is less 
than~$10^{-6}$. This is not a fair comparison, because we are not considering
computational cost whatsoever, but does indicate 
that the~\texttt{bamps} data is superior in this case. Therefore one 
expects that as more resolution is added to the BAM grid the 
result would converge to the~\texttt{bamps} result. Taking data that 
is stronger, for example~$A=5$, the method 
of~\cite{HilBauWey13} failed as an incoming pulse in the lapse 
became evermore sharp, resulting in what seemed to be a coordinate 
singularity. This would be acceptable if an apparent horizon could
be found before the code crashed, but this was not the case.
Going to higher amplitudes still, similar failures occurred, 
and the conclusion was drawn that moving puncture coordinates 
were not suitable for managing this initial data. We have 
already seen in our comparison with~\cite{Rin06} that using~\texttt{bamps}
the~$A=5$ data can be classified supercritical, and we did not
see any sign of a coordinate singularity before an apparent 
horizon was discovered despite the two lapse functions appearing 
qualitatively similar at the beginning. 

\subsection{Comparison with Sorkin} 

Having collected a bank of evidence that our numerical results are
correct we now compare with~\cite{Sor10}. We demonstrate firstly that
we can obtain qualitatively the same type of behavior described
therein, namely that the peak of the Kretschmann scalar
appears away from the symmetry axis. This we achieve however just by
evolving supercritical Brill wave data and changing the gauge source
function we use to evolve it. We find that the position of the peak of
the Kretschmann scalar can be controlled by the choice of gauge source function.
This can be understood geometrically. Secondly, by locating an apparent
horizon in the time development, we demonstrate explicitly that the
amplitude~$A=6.073$, evolved and classified in~\cite{Sor10} as subcritical,
is in fact supercritical. We looked at evolutions of this data at several
resolutions, both of the numerical spacetime and the apparent horizon
computed on top of the data, and find that the outcome is robust.

\paragraph*{Simulation setup:} The simulations of this subsection 
have been carried out on a cubed-ball grid, as described 
in~\cite{HilWeyBru15,Wey14}, with the following setup. The inner 
cube extends from~$r_{\text{cu}}=0.5$ to~$-r_{\text{cu}}$ and is divided 
into~$\mathcal{N}_{\text{cu}}=9$ subpatches with~$N_{\text{cu}}=21$ 
gridpoints in each dimension. For the transition shell from~$r_{\text{cu}}$ 
to~$r_{\text{cs}}=1.0$ we use only one shell~$\mathcal{N}_{\text{cs}}=1$ 
with~$N_{\text{cs}}=35$ points in the radial direction. From here we 
go to the outer boundary at~$r_{\text{ss}}=12$ using~$\mathcal{N}_{\text{ss}}=22$ 
outer shells with also~$N_{\text{ss}}=35$ radial collocation points.
These tests were carried out in 3d using the octant symmetry mode 
of~\texttt{bamps}. 

\begin{figure*}[t]
\centering
\includegraphics[]{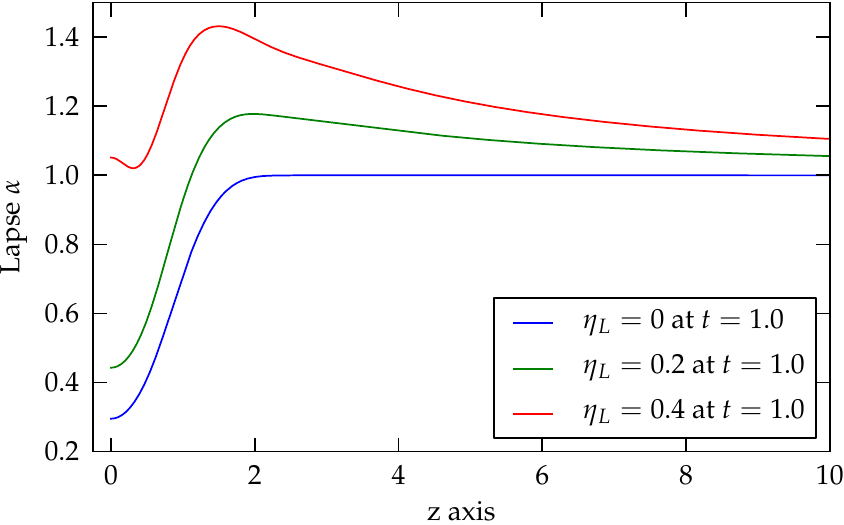}
\includegraphics[]{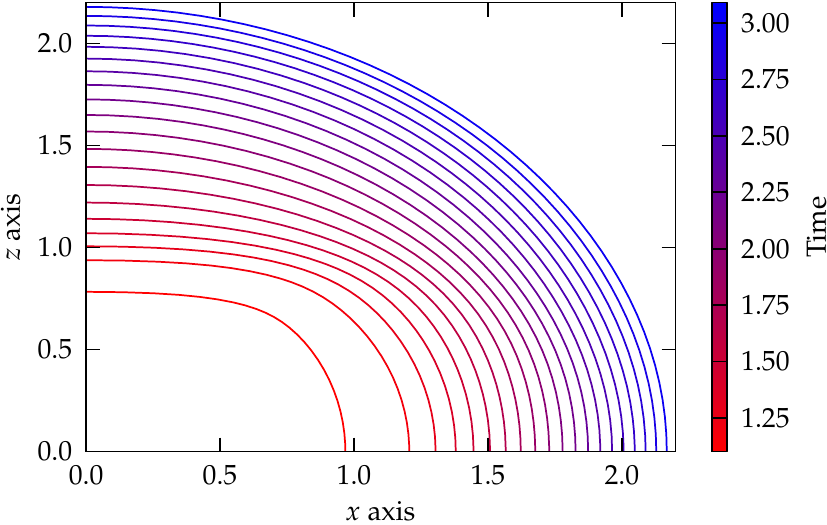}
\caption[]{In the left panel the lapse for the centered~$A=6.073$ 
Brill wave is plotted with the three generalized harmonic 
gauges~$\eta_L=0,0.2$ and~$\eta_L=0.4$ at coordinate 
time~$t=1.0$. The more ``singularity avoiding'' the gauge choice, 
the smaller the lapse becomes around the origin. In the right 
hand panel the apparent horizons Brill wave initial data is 
plotted. The initial horizon mass is~$M_H=0.84$ and has increased 
to around~$M_H=0.90$ before the code fails. These values are to be 
compared with the ADM mass, $M_{\textrm{ADM}}=1.02$.}\label{fig:Sor_AH}
\end{figure*}

\paragraph*{Position of the peak curvature:} Consider a collapse 
spacetime with the standard causal structure. Different foliations of
the spacetime, in which the time coordinate tends to tick more or less
slowly in a region of high curvature depending on some {\it singularity
avoidance} parameter will have different profiles in a spacetime diagram.
If we are given a patch of this spacetime up to a finite time coordinate,
as in a numerical relativity simulation, the specific observer that
encounters the largest curvature before the code crashes depends, among
other things, on the singularity avoidance parameter. In our case such a
parameter is given by~$\eta_L$. We performed evolutions of a centered Brill
wave with~$A=6.073$. This amplitude is shown to be supercritical in the next
paragraph. We evolved with fixed~$\eta_S=6,p=1$ and one of~$\eta_L=0,0.2$ 
or~$\eta_L=0.4$. In the left plot of Fig.~\ref{fig:Sor_AH} we show 
the lapse in the evolutions at coordinate times~$t=1$, which 
demonstrate the effect of the singularity avoidance 
parameter~$\eta_L$. In the pure harmonic slicing case~$\eta_L=0$ we 
have the strongest singularity avoidance, and find that the peak 
of the Kretschmann scalar appears at~$\rho=0.88$, where it simply grows 
until the numerics fail. Increasing the parameter to~$\eta_L=0.2,0.4$ we 
find that the peak of the Kretschmann appears at a coordinate 
radius of~$\rho=0.73$ and~$\rho=0.64$ respectively, before the 
numerics fail.

\paragraph*{Apparent horizon formation:} We evolved the same 
centered~$A=6.073$ data with different resolutions, fixing the 
gauge parameters~$\eta_S=0.4,p=1$ and~$\eta_S=6$, so the largest 
choice of~$\eta_L$ from above. We find rapid convergence of the 
constraints. In particular we used our standard cubed-sphere 
setup with~$N=21^3$, $25^3$ and $29^3$ points per subpatch, and find 
that, for example, the maximum of the~$C_x$ component of the 
Harmonic constraints along the x-axis 
are approximately~$ 5\times 10^{-5},9\times 10^{-6}$ and~$8\times 10^{-7}$ 
respectively at~$t\approx1.25$. We then searched for apparent 
horizons using the method described in~\cite{HilWeyBru15,Wey14}. 
We first find an apparent horizon at around~$t=1$. On a fixed 
numerical spacetime data set we find perfect fourth order convergence 
in the apparent horizon data consistent with the Runge-Kutta method 
employed. Comparing the apparent horizons 
discovered on the different data we find perfect qualitative 
agreement. Furthermore we see behavior consistent with rapid 
convergence when evaluating the differences between the discovered 
horizons. The apparent horizons of different time slices are plotted 
in Fig.~\ref{fig:Sor_AH}, where the coordinate expansion of the 
horizon can clearly be seen. For comparison we again evolved the 
same initial data inside the BAM finite differencing code, but were 
unable with our current setup to find apparent horizons from 
that data. One issue is that the constraint violation is 
many orders of magnitude greater in the finite differencing 
code. Qualitatively however we find good agreement in the 
evolution, at least initially. 

\paragraph*{Summary:} We are unable to reproduce the results
of~\cite{Sor10}, despite using, to the best of our knowledge,
identical initial data and gauge. In fact our results appear to
contradict the earlier study. The reason for the disagreement is
presently not clear. It is possible that an apparent horizon search
on Sorkin's older finite differencing data was too challenging because of
numerical error, or perhaps even that the numerical dissipation was
sufficient to let these strong data spuriously settle down to
flat-space, although the latter does not seem likely. The power-law
scaling obtained in~\cite{Sor10} in the rapid oscillations of the
curvature is nevertheless interesting, and would be good to properly
understand in the future. Given our earlier code
validation~\cite{HilWeyBru15} and the literature comparison in
section~\ref{subsection:comparison} however, we have no reason to
doubt the~\texttt{bamps} results, so for now we move on.

\subsection{Discussion}\label{subsection:Discussion}

\paragraph*{Critical collapse of the scalar-field:} Before moving
to harder experiments, consider the case of the minimally coupled
massless scalar field. Working in spherical symmetry, evolving
families of initial data with strength parameter~$A$, Choptuik
found~\cite{Cho93} compelling numerical evidence for the existence
of a critical solution at~$A=A_\star$ serving as the boundary
between dispersion and collapse. The appearance of critical
phenomena with~$A$ in a neighborhood of~$A_\star$ was neatly
explained by the conjecture that the critical solution is an
attractor of co-dimension one in phase space. In other words in
a neighborhood of the critical solution there should to be just
one growing mode. Working in perturbation theory around the
critical solution, but crucially allowing for aspherical mode
perturbations, Gundlach and Mart\'in-G\'arc\'ia found strong
numerical evidence for this conjecture~\cite{GarGun98}. They
found that the most slowly decaying mode came with an eigenvalue
of~$-\lambda_1\approx 0.02$ associated with a~$Y_{20}$ spherical
harmonic. For this matter model the single unstable mode has
eigenvalue~$\lambda_0\approx 2.7$. Because of the exponential
decay of the former it was conjectured that the qualitative
picture obtained in spherical symmetry using the full Einstein
equations would not be altered for generic initial data close
to criticality. This picture, roughly speaking, says that in a
neighborhood of critical collapse the fields should strongly
interact in a confined region for a finite, but ever longer time
as the critical solution is approached, and ultimately either
collapse or disperse. Interestingly for the current study,
Choptuik and collaborators~\cite{ChoHirLie03} then studied
axisymmetric configurations and found evidence of a second growing
mode associated with a~$Y_{20}$ spherical harmonic, causing the strong
field solution to bifurcate into {\it two} strongly interacting regions,
in apparent contradiction with the earlier perturbative result.

\paragraph*{Issues with the apparent horizon as a diagnostic:} The
apparent horizon is used for classification of the spacetime as sub or
supercritical. In earlier work the apparent horizon mass has
also been used to indicate power-law scaling in the critical regime.
Therefore it is worth noting explicitly the weaknesses of this approach.
Firstly, the appearance of an apparent horizon guarantees the existence
of an event horizon only in strongly asymptotically predictable spacetimes,
which arise from generic asymptotically flat initial data only if the
weak-cosmic censorship conjecture holds~\cite{HawEll73,Wal84}. This
fact makes our approach blind to violations of weak-cosmic censorship.
It is also possible that no apparent horizon appears in our particular
foliation, even if there is a black hole region~\cite{WalIye91}.
Secondly the foliation dependence of the apparent horizon makes it
difficult to trust the black hole masses obtained from the horizon
area, particularly away from spherical symmetry. We may try to diagnose
power-law scaling by looking at horizon mass, but the foliation could be
such that the apparent horizon always forms with large area, and we have
to choose {\it when} to evaluate the mass for each member of the one-parameter
family. In fact in preliminary experiments we found that when approaching
the critical regime, behavior resembling power-law scaling in the initial
horizon masses could appear with a particular choice of gauge source
function, but then completely disappear once we altered the choice to
avoid coordinate singularities. These difficulties make us strongly prefer
to study the subcritical approach to black hole formation, because
there we can, as in~\cite{Sor10}, unambiguously consider the spacetime
maximum of the Kretschmann scalar, provided that we are able to
evolve the spacetime long enough to be confident that it is really
subcritical.

\subsection{Towards the critical regime}

\begin{table}[t]
  \centering
  \begin{ruledtabular}
  \caption[]{\label{tab:crit_grids}
In the upper table we summarize the runs obtained in each sample,
with~$\Delta A=(0.1)^i$, and afterwards in the bisection search. $A$ is the
strength parameter of the wave~\eqref{eqn:brill_seed}, $M_{\textrm{ADM}}$ the
ADM mass of the initial data set, $t_{\text{AH}}$ the coordinate time at which
an apparent horizon was first discovered, $M_H$ the horizon mass at first
and at the end of the simulation, $t_c$ the coordinate time that the
code crashed. The final column summarizes changes to the setup not 
described in the main text. We denote the strength parameters of our
current bounding runs in bold. In the lower table the various cubed-ball
grids are specified using the notation of~\cite{HilWeyBru15}. The
grid parameters are~$r_{\text{cu}}$ the extent of the central
cube,~$r_{\text{cs}}$ the radius at which the transition shell ends, 
$r_{\text{ss}}$ the coordinate position of the outer
boundary,~$\mathcal{N}_{\text{cu}}$ the number of subpatches per
dimension in the central cube, $\mathcal{N}_{\text{cs}}$,
$\mathcal{N}_{\text{ss}}$ the number of radial subpatches within the
transition and outer shells respectively
and~$N=N_{\text{cu}}=N_{\text{cs}}=N_{\text{ss}}$ the number of collocation 
points per dimension per subpatch in our lowest resolution runs on this
grid.
}
  \begin{tabular}{clccccc}
      Sweep & $A$ & $M_{\textrm{ADM}}$ & $t_{\text{AH}}$ & $M_H$ & $t_{\textrm{c}}$ & Changes\\
    \hline
      $i=1$  & $4.7$ & $0.622$ & $15.0$ & $0.27$/$0.30$ & $16.6$ & \\
         & $4.6$ & $0.597$ &  $\text{\sffamily X}$ &  $\text{\sffamily X}$ &  $\text{\sffamily X}$ &\\
    \hline
      $i=2$  & $4.70$ & $0.622$ & $15.0$ & $0.27$/$0.30$ &  $16.6$ & \\
         & $4.69$ & $0.619$ & $\text{\sffamily X}$ & $\text{\sffamily X}$ & $\text{\sffamily X}$ & $p=0$\\
    \hline
      $i=3$  & $4.697$ & $0.621$ & $16.4$ & $0.08$/$0.08$ & $16.5$ & $p=0.5$\\
         & $4.696$ & $0.621$ & $\text{\sffamily X}$ & $\text{\sffamily X}$ & $\text{\sffamily X}$ & $p=0.5$\\
    \hline
      Bisect  & & & & & &\\
    \hline
      1 & $4.6965$ & $0.621$ & $\text{\sffamily X}$ & $\text{\sffamily X}$ & $\text{\sffamily X}$ & $p=0.5$\\
      2 & $4.69675$ & $0.621$ & $17.5$ & $0.03$/$0.03$ & $17.7$ & $p=0.5$\\
      3 & $4.696625$ & $0.621$ & $\text{\sffamily X}$ &  $\text{\sffamily X}$ & $\text{\sffamily X}$ & $p=0.5$\\
      4 & $\mathbf{4.6966875}$ & $0.621$ & $\text{\sffamily X}$ & $\text{\sffamily X}$ & $\text{\sffamily X}$ & $p=0.5$\\
      5 & $4.69671875$ & $0.621$ & $17.9$ & $0.03$/$0.05$ & $18.6$ & $p=0.5$\\
      6 & $\mathbf{4.696703125}$ & $0.621$ & $18.7$ & $0.03$/$0.06$ & $19.0$ & $p=0.5$
  \end{tabular}
  \vspace{0.25cm}
  \begin{tabular}{cccccccc}
   Grid & $r_{\text{cu}}$ & $r_{\text{cs}}$ & $r_{\text{ss}}$ & $\mathcal{N}_{\text{cu}}$ & $\mathcal{N}_{\text{cs}}$ & $\mathcal{N}_{\text{ss}}$ & $N$\\
   \hline
   G0 & $1.0$ & $7.0$ & $20.0$ & $33$ & $11$ & $13$ & $31$ \\
   G1 & $2.5$ & $10.5$ & $21.5$ & $63$ & $11$ & $13$ & $31$ \\
   G2 & $2.5$ & $10.5$ & $21.5$ & $95$ & $11$ & $13$ & $31$ \\
   G3 & $3.0$ & $12.0$ & $21.0$ & $199$ & $18$ & $13$ & $19$ \\
  \end{tabular}
  \end{ruledtabular}
\end{table}

\begin{figure*}[t!]
\centering
\includegraphics[width=\columnwidth]{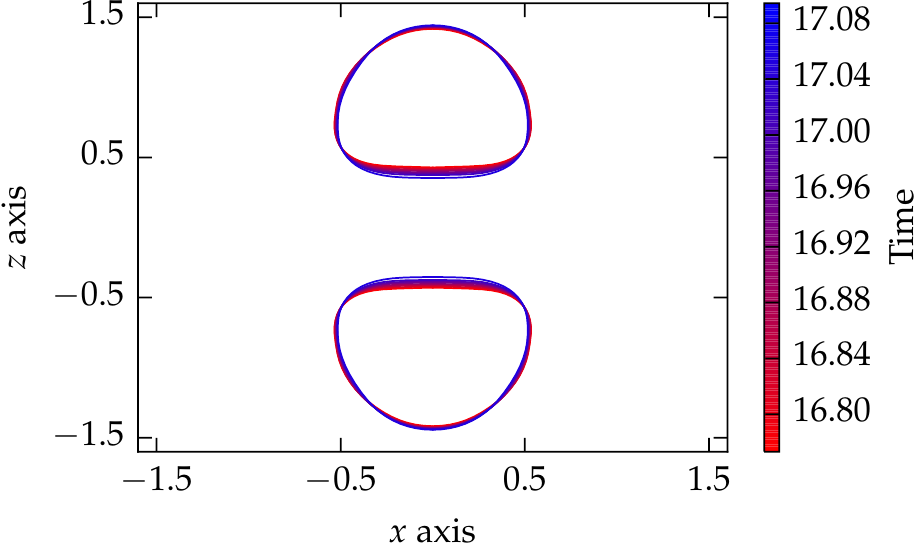}
\includegraphics[width=\columnwidth]{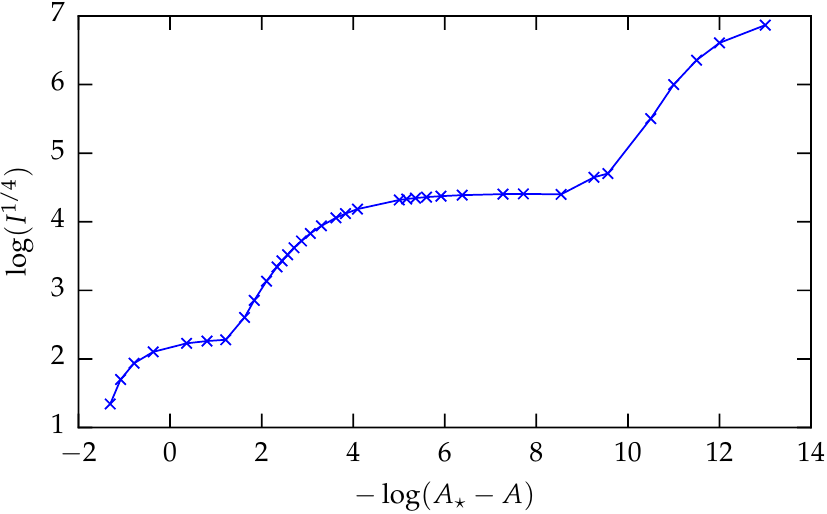}
\caption[]{In the left panel the apparent horizons at different times
in~$A=4.698$ centered Brill wave initial data, as obtained in sweep~$3$
with~$\Delta A=(0.1)^3$, are plotted. Evidently two apparent horizons
appear in the data, each around the observed peaks in the Kretschmann
scalar at~$z=\pm z_{\text{peak}}$, indicating the likelihood that the 
family results in head-on binary black hole spacetime near the
critical amplitude. This behavior is robust in that in weaker
supercritical data that we can successfully classify, we always
find such horizons. In the right hand panel we plot the logarithm
of the absolute value of the Kretschmann scalar against~$-\log(\As-A)$ 
taking~$\As=4.6966953125$ as the critical amplitude.
The result can be well-fitted by a straight-line with
gradient~$\sim0.37$ plus a function of period~$\sim8$. This is
indicative of critical behavior~\cite{GarDun98}, but since we see
only one full period, starting from around~$-\log(\As-A)=2$, we
do not consider the result conclusive. 
\label{fig:AH_off_Kret_Scale}}
\end{figure*}

\paragraph*{Search strategy:} Our previous tests demonstrate 
that apparent horizon formation first occurs between~$A=4$ and~$A=5$.
We therefore searched for a critical amplitude~$\As$ in this range.
Running exclusively in cartoon mode, we started with that
bracketing and within it sampled~$A$ at~$10$ equally spaced values, thus
obtaining a new bracket~$10$ times smaller. We did this in three times,
with~$\Delta A=(0.1)^i$ for~$i=1,2,3$. At each level, once the
critical amplitude is bracketed we increase resolution on the
bracketing amplitudes to check convergence and be sure that we truly
have bracketed~$\As$. After the third such sweep we went to a straightforward
bisection search, increasing resolution, adjusting grids and gauge
sources as seemed appropriate. Our current best bound for the critical
point is that~$\As\in[A_L,A_U]=[4.6966875,4.696703125]$, an interval of
width~$\sim 1.6\times10^{-5}$. Afterwards we performed additional runs
away from the critical point to help understand the behavior of the
Kretschmann scalar and the initial apparent horizon masses as a
function of~$A$. We started with the gauge source parameters
\begin{align}
\eta_L=0.4\,\alpha^{-2}\,, \quad p=1\,, \quad \eta_S=6\,, 
\end{align}
and take our defaults for the GHG formulation settled 
on in~\cite{HilWeyBru15,Wey14}. The parameters of our base grid (G0
in Tab.~\ref{tab:crit_grids}) result in a total of~$1105$ subpatches
in cartoon mode. Spreading this over the maximum number of cores
possible on {\it SuperMUC} (i.e.\ one subpatch per core), 
the code computes at around~$3$ time
units per hour at the lowest resolution. Modifications to the gauge
and grid are summarized in Tab.~\ref{tab:crit_grids}, which gives
the results both of the bracketing for each sweep and in the
bisection search. Grids G1, G2 and G3 consist of~$2560,4608$
and~$16200$ subpatches, respectively. The ADM masses of each initial
data set are also given in the table, although one should be careful
to remember that we impose boundary conditions at a finite coordinate
radius, which could affect the dynamics of the evolution, and
furthermore makes the interpretation of~$M_{\textrm{ADM}}$ non-trivial.

\paragraph*{Termination of search:} It may be possible to push to a
better bound by brute force with the current method, but we stopped our
bisection search at the range~$\As\in[A_L,A_U]$ as we prefer to conserve
resources to attack alternative initial data. A main issue preventing us
from going further~{\it economically} is the lack of mesh-refinement
inside~\texttt{bamps}. The implementation of a true pseudospectral
adaptive mesh-refinement algorithm is a major undertaking due to its
complexity, and because efficient parallelization then becomes challenging.
Another issue is that our initial data solver can only solve the Hamiltonian
constraint down to around the~$10^{-10}$ level, which is presumably
caused by the use of irregular coordinates, the Chebyschev-Fourier-Fourier
discretization and simply machine precision~\cite{Wey14}, but this level
of error is certainly not the leading order in our present simulations.
It is curious that in studies with various matter models it has been possible
to fine-tune to much higher accuracy, even in cases where the basic accuracy
of the numerical method is much lower than in ours. This can often be achieved
by keeping the numerical resolution and everything else, except the amplitude
fixed. However, the fine-tuning error in~$\As$ for a given resolution can
be much smaller than the drift towards convergence when increasing the
resolution. As we explain below, we believe that in the present simulations
the main issue preventing us from going further in the fine-tuning of~$\As$
is related to gauge choice.

\paragraph*{Description of dynamics:} The basic dynamics from
each of the initial data are initially rather similar. At first
a pulse in the Kretschmann scalar propagates out from the origin
predominantly in the~$\rho$ direction. The pulse then propagates
more slowly, eventually turning around and traveling towards the
origin. As it propagates in, the pulse is smeared out parallel
to the $z$-axis. As the pulse hits the axis, there is a rapid
growth resulting in a maximum at some~$\pm z_{\text{peak}}\ne0$. In
the first sweep, for~$A\geq 4.8$ data an apparent horizon is
found around or just after the time of this growth. In the~$A=4.7$
run, this peak in the Kretschmann occurs at~$z_{\text{peak}}=1.25$,
with a value of~$5.5\times10^{7}$. The feature then starts to
propagate away, predominantly in the~$\rho$-direction, and no
apparent horizon is found until later. Instead the evolution
continues until a second large peak appears in the
Kretschmann scalar around~$z_{\text{peak}}\simeq1$ on the symmetry
axis, as the wave content leftover from the first big peak again
crashes onto the axis. An apparent horizon is found shortly
afterwards and consistently until the evolution fails at~$t\simeq16.6$,
as the Kretschmann scalar starts to grow ever more rapidly around
these peaks. In the second sweep we switch the slicing 
condition to~$p=0$ to avoid spikes appearing in the lapse. The
largest amplitude of this sweep, $A=4.69$, is subcritical. The peak
of the Kretschmann in this run is around~$3.25\times10^{7}$ and
appears at~$z_{\text{peak}}=1.34$; a movie of the dynamics can be
found online~\cite{HilWebsite}. As we go closer to the critical
point we see the appearance of yet-more spikes, which then
propagate up and down the axis. This behavior is discussed in
more detail in the following paragraphs.

\paragraph*{Disjoint apparent horizons:} Starting with the third
sweep,~$\Delta A=(0.1)^3$, we find in supercritical data {\it two disjoint
apparent horizons}, centered roughly around the position of the second
large feature growing on the axis at~$\pm z_{\text{peak}}$. An example is shown in
Fig.~\ref{fig:AH_off_Kret_Scale}. In other words, close to the critical
point these initial data produce what seem to be axisymmetric binary
black hole spacetimes! Obviously in the
case of gravitational waves spherical decay is impossible, so this
bifurcation is perhaps the generic near-critical behavior. Evidence for this
could be sought by evolving different families of data. We expect that the
reflection symmetry about the x-axis plays a role here in the outcome
however. For generic axisymmetric supercritical data, with one loosely
defined strong-field region, one might expect that the bifurcation happens
but that apparent horizon formation appears on only one side. This result
means that with our current setup we are not able to evolve to a final end-state,
as \texttt{bamps} does not have a moving-excision setup or the control
mechanism of SpEC~\cite{SchGieHem14,HemSchKid13}. Within the lifetime of our
simulations we do not find a common horizon surrounding the disjoint MOTS.
But the lifetime of the simulations after horizon formation is short,
so this to be anticipated. To be sure that these spacetimes really
do contain two black holes it will be necessary to search for an event horizon,
but the short lifetime prohibits this also. Nevertheless the fact that as we
get closer to the critical solution, the initial apparent horizon size gets
smaller, whilst the coordinate distance between the horizons remains roughly
constant hints that the spacetimes do contain two black holes. In the context
of critical collapse we are predominantly interested in the strong-field region
near to the critical threshold, so we continue with the search, focusing
primarily on the subcritical regime. The detailed study of supercritical data
is left for future work. 

\paragraph*{Scaling of the Kretschmann scalar:} According 
to~\cite{GarDun98}, if critical phenomena are present during 
gravitational collapse, then one should see power-law
scaling of curvature invariants in the subcritical regime~$A\lesssim\As$. 
Since we are working in vacuum any scalar built from the Ricci
curvature is unavailable so, as in~\cite{Sor10}, we focus on the
Kretschmann scalar. In the right hand panel of Fig.~\ref{fig:AH_off_Kret_Scale}
we plot the maximum value of the Kretschmann scalar in the spacetimes as a
function of~$A-\As$ in a log-log plot. There is a plateau in the maximum
before it starts to increase rapidly in~$A-\As$. The rapid increase
occurs as a later implosion of the wave onto the axis starts to
dominate over the previous implosion. The resulting curve can be
fit as,
\begin{align}
\log(I_{\textrm{max}}^{1/4})\simeq-\gamma\log(\As-A)+\Psi[\log(\As-A)]\,,
\label{eqn:scaling}
\end{align}
with~$\gamma\sim0.37$ and~$\Psi$ of period around~$8$. Over one period
the maximum in the Kretschmann scalar increases by a factor of~$e^{4\Delta}$,
with~$\Delta\sim3$. The first of these numbers agrees with that obtained
in~\cite{AbrEva93}, whilst our value of~$\Delta$ does not. It is not yet
clear how seriously these numbers should be taken because so far we
observe only one full period in the Kretschmann scalar. Therefore we
postpone the assignment of error bars and for now simply advise caution
against over-interpretation of the finding.

\paragraph*{Comparison with the scalar-field:} Our results are 
reminiscent of the bifurcation of the scalar field~\cite{ChoHirLie03} discussed
earlier in section~\ref{subsection:Discussion}. It is possible that there
is a direct relationship; our results indicate that in vacuum axisymmetry,
near the critical solution, decay proceeds by the aforementioned bifurcation.
On the other hand, we know empirically that in spherical symmetry dispersion
of the scalar field is determined by a single unstable spherical mode. Take
a spacetime with a single strong-field region with scalar field and gravitational
wave content. Imagine fixing, in some sense, the ratio of
gravitational wave and scalar field content, and heading towards the
threshold of black hole formation. By continuity, we expect the critical
solution to interpolate between the two scenarios of bifurcation (driven by
the gravitational wave content) and spherical decay (driven by the
scalar-field content) as the ratio of the two is adjusted. This idea
is compatible with~\cite{ChoHirLie03}, in which gravitational wave content
is added to the initial data by placing axisymmetric perturbations on the
metric. It is also compatible with the {\it perturbative} results
of~\cite{GarGun98}. By construction the Choptuik solution is absent of
gravitational waves, so a linear analysis could not spot
a complicated nonlinear admixture of the decay mechanisms.
Alternatively, it may not matter whether the strong-field is
formed by a gravitational wave or other source. Comparing spherically
symmetric with axisymmetric evolutions, the main difference is that it is
possible to form multiple centers of collapse in axisymmetry. With
sufficiently large asphericity this could be generic.

\paragraph*{Wishlist for future work:} Evidence for the above suggestion 
could be sought in several obvious ways. First, evolution of different families
of axisymmetric vacuum data must be performed to see whether or not the bifurcation
behavior really is generic. Next, it would be good to compute accurately the value
of~$\Delta$ in vacuum (assuming that the tentative behavior persists) and to
compare with the value obtained by~\cite{ChoHirLie03} as the critical solution
is deformed by gravitational wave content. At first glance our
value~$\Delta\sim3$ appears consistent with that of~\cite{ChoHirLie03}, but at
this stage nothing is certain. Another possibility is to work in second order
perturbation theory about the Choptuik critical solution and to look for
evidence of the bifurcation behavior. Finally, more results for the critical
axisymmetric scalar field are also highly desirable.

\begin{figure}[t!]
\centering
\includegraphics[width=\columnwidth]{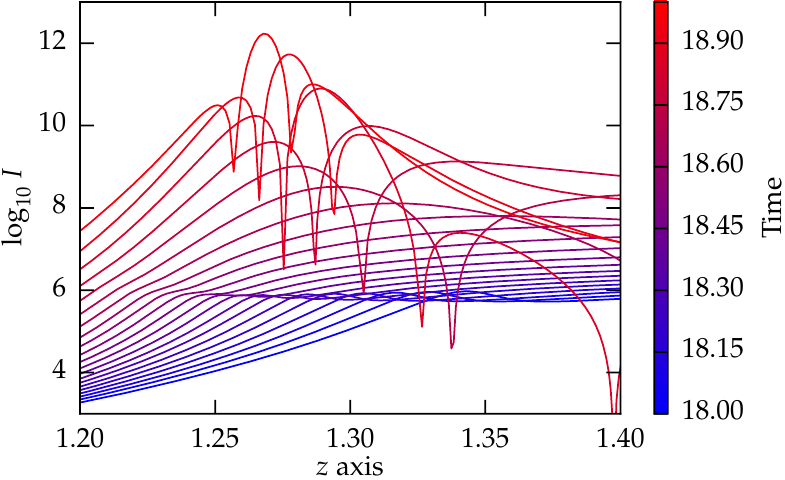}
\caption[]{Here we plot the logarithm of the Kretschmann scalar
along the symmetry axis around the times when the largest spikes
appear in the~$A=4.6966953125$ experiment. Note that we have not
classified this spacetime as sub or supercritical because with
our current setup doing so will be very expensive to do so with
confidence. We believe that the crash is caused by a coordinate
singularity however, which forms as the biggest peak
dissipates.\label{fig:Kret_Spike}}
\end{figure}

\paragraph*{Spikes in the Kretschmann Scalar and code failure:} For
our final test, which we can not yet classify, close to the critical point
we evolved initial data with amplitude~$A=4.6966953125$. If this experiment
were successful it would correspond to the next bisection step.
We find that there are a sequence of large spikes on the symmetry axis as
the gravitational wave implodes, then propagates up and down the symmetry
axis before imploding once more. Each of the large spikes is finer and
therefore requires more resolution for accuracy. It is tempting to label the
sequence of strong oscillations `echoes', but again, perhaps
because of a suboptimal gauge, we can not quantify this
claim and therefore resist. Fig.~\ref{fig:Kret_Spike} shows the
run-up to and the evolution of the final spike before the code
crashes. In practice the numerics fail not as such a spike forms
but rather as it dissipates away. As this happens we see along the~$z$-axis
that a sharp feature suddenly forms in the metric component~$g_{zz}$
and causes the code to crash. The difficulty is in classifying the spacetime
rather than the code crash per se, as do not find an apparent
horizon in the data before the crash. The cause of the feature is
unclear, but possible candidates are simple numerical error, the
formation of an apparent horizon that the present method is unable to
unveil, the formation of a coordinate singularity, or even the
seemingly unlikely formation of a naked singularity. Increasing
resolution very substantially hardly affects the appearance of
the feature or crash-time of the code, so in this case it is
doubtful that even mesh-refinement could address the problem
directly. The main suspect is therefore the formation of a
coordinate singularity. This view is further enforced by the fact
that coordinate problems, although of a different specific form, also
occurred in simulations with the qualitatively similar 1+log slicing
using the BSSNOK formulation~\cite{San06,HilBauWey13}. To investigate
this we have evolved with different gauge source parameters. Informed
by earlier experience we increased~$\eta_S$. This however has the
unfortunate side-effect of allowing the strong-field region to bleed
out from the central cube into the transition shell where we have lower
resolution, and so results in other problems. Going to slightly lower
wave amplitudes, the sudden spike in~$g_{zz}$ persists even with a large
value of~$\eta_S$, albeit at a later coordinate time. Ultimately a
radical change of coordinates may be needed. Addressing the problem
with an improved continuum formulation and numerical method is a
priority.

\section{Conclusions}\label{section:Conclusions}

We have continued our study of gravitational
waves in the regime separating dispersion from black hole
formation. To maximize overlap with earlier results we focused 
exclusively on Brill waves with the seed function~\eqref{eqn:brill_seed},
and evolved only prolate~($\sigma_\rho=\sigma_z=1$), geometrically
prolate~($A>0$) centered~($\rho_0=z_0=0$) data. Our main findings are
first that, while our results are in agreement with several other
publications, we are unable to reproduce  those of~\cite{Sor10},
despite performing evolutions of the same initial data with the
same gauge conditions. In particular we unambiguously find apparent
horizons in data classified there as subcritical. The reason for
this difference is not clear. Moving closer to the
threshold of black hole formation, surprisingly, we find that
two disjoint apparent horizons are found centered around some
non-zero~$z_0$ on the symmetry axis, indicating the likelihood
that the Brill wave collapses to form a head-on collision of two
black holes. The fact of, and time-scale for the merger
of these horizons is to be determined. Finally we have bounded
the critical amplitude within a range of about~$10^{-5}$. This is an
improvement over the previous bound by some orders of magnitude.
We see evidence of power-law scaling, since the maximum
of the Kretschmann scalar is well described by the form~\eqref{eqn:scaling},
as expected if critical phenomena are present~\cite{GarDun98}. The power-law
exponent, at least, appears consistent with the value of Abrahams
and Evans~\cite{AbrEva93}, but our tentative value of~$\Delta\sim3$ is very
different from theirs~$\Delta\simeq0.6$. On the other hand our 
value of~$\Delta$ is compatible with that of the mixed axisymmetric
gravitational wave and scalar-field data~\cite{ChoHirLie03}. Since we only
see one period of the wiggle however, we must warn against premature
jubilation. The wiggle could disappear, or the period of subsequent wiggles
may differ substantially with more tuning of~$A$, so further work is 
definitely needed. Closer to the critical point we find more and more extreme
behavior in the Kretschmann scalar. Particularly interesting are the ever-finer
spikes that rapidly form on the symmetry axis. Superficially this even seems
evocative of BKL type behavior.

Close to the critical point our current method suffers from larger
errors, particularly in the form of constraint violation around the
spikes and, we suspect, because coordinate singularities form. Evidently
there is still much to understand. The next steps will include looking at
different initial data, including Brill waves with~$A<0$, prolate and
off-centered seed functions, along with the Teukolsky waves
of~\cite{AbrEva93}, which suit better the original spirit of~\cite{Cho93}
since they consist of incoming colliding waves. In the future we
furthermore hope that the combination of mesh-refinement and the use of
the dual-foliation~\cite{Hil15,HilHarBug16} approach will help to allay
our current difficulties. 

\acknowledgments

We are grateful to David Garfinkle, Sascha Husa, 
Anton Khirnov, Tom\'a\v{s} Ledvinka and Hannes R\"uter for interesting
discussions, and especially to Carsten Gundlach for helpful comments on
the manuscript. This work was supported in part by the FCT (Portugal) IF
Program~IF/00577/2015, the Deutsche Forschungsgemeinschaft (DFG)
through its Transregional Center SFB/TR7 ``Gravitational Wave Astronomy'',
by the DFG Research Training Group 1523/2 ``Quantum and Gravitational
Fields'', and the Graduierten-Akademie Jena. Computations were performed
primarily at the LRZ (Munich), but also on the Jena ARA cluster. 

\begin{appendix}

\end{appendix}

\bibliographystyle{unsrt}
\bibliography{Brill.bbl}{}

\end{document}